\documentclass[aps,prd,twocolumn]{revtex4}

\usepackage{graphicx}  % standard LaTeX graphics tool

 \usepackage{amssymb}
 
\begin{document}

\title{Power Transfer in  Non-linear Gravitational Clustering and Asymptotic Universality }

\author{T.~Padmanabhan$^a$, Suryadeep Ray$^b$}

\email[$^a$]{Email: nabhan@iucaa.ernet.in}
\email[$^b$]{Email: surya@iucaa.ernet.in}

\affiliation{IUCAA,
Post Bag 4, Ganeshkhind, Pune - 411 007, India.\\
}

%%upright Greek letters (example below: upright "mu")
\newcommand{\greeksym}[1]{{\usefont{U}{psy}{m}{n}#1}}
\newcommand{\umu}{\mbox{\greeksym{m}}}
\newcommand{\udelta}{\mbox{\greeksym{d}}}
\newcommand{\uDelta}{\mbox{\greeksym{D}}}
\newcommand{\uPi}{\mbox{\greeksym{P}}}

%AUTHOR_STYLES_AND_DEFINITIONS%%%%%%%%%%%%%%%%%%%%%%%%%%%%%%%

  \def\bld#1{{\bf #1 }}
 \def\xb{\bar\xi(a,x)}
 \def\gaprox{\mbox{$\,$ 
\raisebox{0.5ex}{$<$}\hspace{-1.7ex}{\raisebox{-0.5ex}{$\sim$ }}$\,$} }
\def\part#1#2{\frac{\partial #1}{\partial #2}}
\def\pder#1#2{{\partial #1/\partial #2}}

\def\rb{\right)}
\def\lb{\left(}

\def\frab#1#2{\left({#1\over#2}\right)}
\def\fra#1#2{{#1\over#2}}

%%%%%%%%%%%%%%%%%%%%%%%%%%%%%%%%%%%%%%%%%%%%%%%%%%%%%%%%

\begin{abstract}
We study the non-linear gravitational clustering of collisionless  particles in an expanding background
using an integro-differential equation for the gravitational potential. In particular, we address the 
question of how the non-linear mode-mode coupling transfers power from one scale to another in the Fourier space if the initial power spectrum is sharply peaked at a given scale. We show that the dynamical equation allows self similar evolution for the gravitational potential $\phi_{\bf k}(t)$ in Fourier space of the form $\phi_{\bf k}(t) = F(t)D({\bf k}) $ where the function
$F(t)$ satisfies a second order non-linear differential equation. We provide a complete analysis
of the relevant solutions of this equation thereby determining the asymptotic time evolution of the gravitational
potential and density contrast.  The analysis shows that both $F(t) $ and $D({\bf k})$ have well-defined asymptotic forms
indicating that the power transfer leads to a universal power spectrum at late times. The analytic results are
compared with numerical simulations showing good agreement.
\end{abstract}

\maketitle

If 
power is injected at some scale $L$ into an ordinary viscous fluid, it cascades down to smaller scales because of the
non-linear coupling between different modes. The resulting power spectrum, for a wide range of scales, is well approximated by the Kolmogorov spectrum which plays a key useful role in the study of fluid turbulence. It is possible to obtain the form of this spectrum
from fairly simple and general considerations though the actual equations of fluid turbulence are 
intractably complicated. 

In this paper, we address the corresponding question for non-linear gravitational clustering of collisionless particles in an expanding background: If power is injected at a given length scale
very early on, how does the dynamical evolution transfer power to other scales at late times?
In particular, does the non-linear evolution lead to an analogue of Kolmogorov spectrum with some level of universality, in the case
of gravitational interactions?

\begin{figure*}[t]
\includegraphics[scale=0.5]{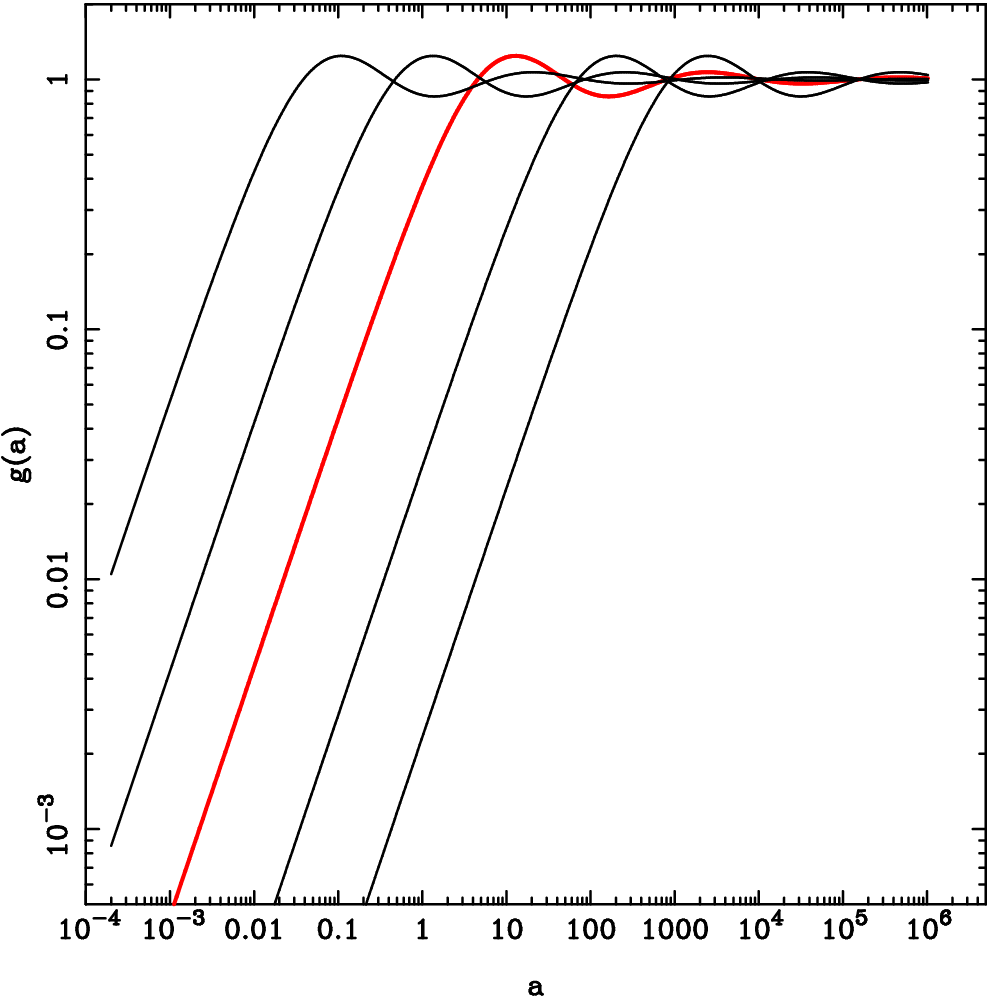}\ \includegraphics[scale=0.5]{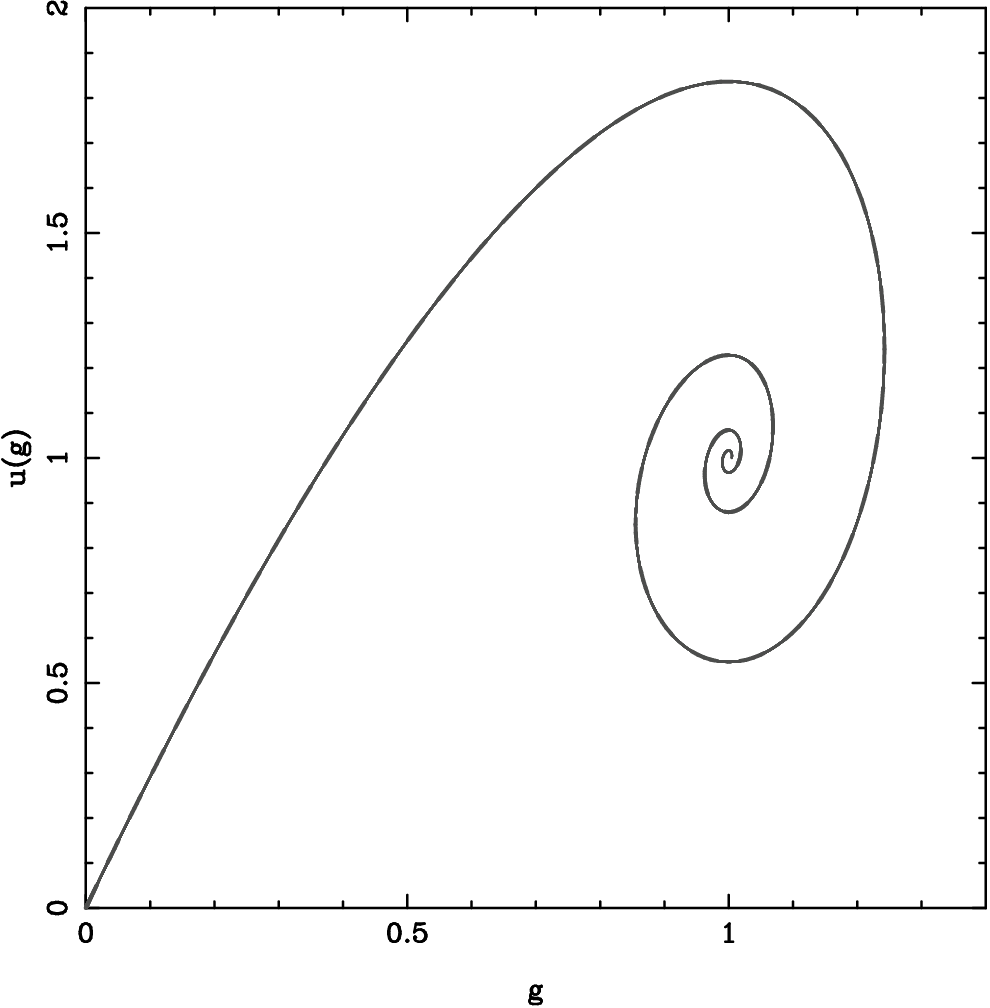}
\caption{The solution to Eq.~(\ref{gofa}) is plotted in both the $g - a$ plane and the $u - g$ plane. The 
function $g(a)$ asymptotically approaches unity with oscillations which are represented by the
spiral in the right panel. The different curves in the  left panel correspond to the rescaling freedom in the 
initial conditions. The entire family of solutions is represented by a single curve
in the $u - g$ plane. One fiducial curve which was used to model the simulation is shown in the
red.}
\end{figure*}

We will show  that the answer is essentially in the affirmative. If power is injected
at a given scale $L = 2\pi/k_0$ then the gravitational clustering transfers the power to both
larger and smaller spatial scales. At large spatial scales the power spectrum goes as $P(k) \propto k^4$
as soon as non-linear coupling becomes important. (This result is known in literature; see for 
example \cite{lssu,gc1}.) More interestingly, the cascading of power to smaller scales leads to a 
\textit{universal pattern} at late times just as in the case of fluid turbulence. By studying the relevant
equations we will show that the gravitational potential has the form $\phi_{\bf k}(t) = F(t)D({\bf k}) $
where $F(t)$ satisfies a non-linear differential equation and $D({\bf k})$ satisfies an integral
equation. We  analyze the relevant equations analytically as well as verify the conclusions
by numerical simulations. The study confirms that non-linear gravitational clustering does
lead to a universal power spectrum at late times if the power is injected at a given scale
initially. 

While the gravitational evolution of collisionless particles in an expanding background is of 
considerable importance in cosmology, it is probably fair to say that our analytic
understanding of this problem is rather patchy. (For a general review of statistical mechanics of gravitating systems,
see \cite{tppr}; previous work on  analytic approximations include Zeldovich-like approximation \cite{za}, perturbative techniques \cite{pi}, and non-linear scaling relations \cite{nsr} among many other approaches.)
In cosmology there is very little  motivation to study the transfer of power by itself
and most of the numerical simulations in the past concentrated on evolving broadband initial power spectrum.

We shall now provide the mathematical details of our formalism.
If ${\bf x}_i(t)$ is the trajectory of the $i-$th particle, then equations for 
 gravitational clustering in an expanding universe in the Newtonian
 limit can be summarized as 
\begin{equation}
\ddot{\bf x}_i + { 2\dot a \over a} \dot{\bf x}_i = - {1 \over a^2} \nabla_{\bf x}
\phi;\quad \nabla_x^2 \phi = 4\pi G a^2 \rho_b \delta  \label{twnine}
\end{equation}
where $\rho_b(t)$ is the smooth background density of matter. Usually one is interested in the evolution of the density contrast $\delta \lb t, \bld x \rb \equiv
[\rho(t,{\bf x}) - \rho_b(t)]/\rho_b(t)$ rather than in the trajectories. Since the density 
contrast in the Fourier space $\delta_{\bf k}(t)$ can be related to the trajectories of the 
particles, one can obtain an equation for $\delta_{\bf k}(t)$ from the above equation.
This equation, in turn, will involve the velocities of the particles and hence will not be
closed. It is however possible to provide a closure condition for this equation using the 
following two facts. 
(a) At any given time $t$, particles which are already part of a bound, virialized cluster do
not contribute significantly to the non-linear terms. 
(b) The velocities of the remaining particles can be very well approximated by the Zeldovich
ansatz which takes the velocities to be proportional to the gradient of the gravitational potential.
Given this ansatz, it is possible to write down a {\it closed} integro-differential equation 
for the evolution of gravitational potential $\phi_{\bld k}(t)$ in the Fourier space. (The details of this derivation
can be found in the companion paper \cite{gc1} and will not be repeated here.)
The final equation is 
\begin{eqnarray}
\ddot \phi_{\bld k} + 4 {\dot a \over a} \dot \phi_{\bld k} &=& -{1 \over 3a^2} \int {d^3 \bld p \over (2 \pi)^3} \phi_{{1 \over 2} \bld k + \bld p} \phi_{{1 \over 2} \bld k - \bld p} {\cal G} (\bld k, \bld p)\nonumber \\
{\cal G} (\bld k, \bld p) &= &{7 \over 8} k^2 + {3 \over 2} p^2 - 5 \lb {\bld k \cdot \bld p\over k}\rb^2 \label{calgxx} 
\end{eqnarray}

This equation governs the dynamical evolution of the system but, of course, is intractably complicated.
In addition to the obvious difficulty of solving an integro differential equation, we also 
need to tackle the issue of incorporating appropriate initial conditions, which will influence the evolution at the early stages. (Some results based on perturbation solution to this equation are discussed
\cite{gc1}.) Our aim is to look for \textit{late time} scale free evolution of the system exploiting the 
fact that the above equation allows self similar solutions of the form 
$\phi_{\bf k}(t) = F(t)D({\bf k}) $. Substituting this ansatz into Eq.~(\ref{calgxx}) we obtain
two separate equations for $F(t)$ and $D({\bf k})$. It is also convenient at this stage
to use the expansion factor $a(t) = (t/t_0)^{2/3}$ of the matter dominated universe
as the independent variable rather than the cosmic time $t$. Then simple algebra shows that
the governing equations are 
\begin{equation}
a \frac{d^2 F}{da^2} + \frac{7}{2} \frac{dF}{da} = - F^2
\label{tevl}
\end{equation}
and 
\begin{equation}
 H_0^2 D_{\bf k} =\frac{1}{3}\int \frac{d^3{\bf p}}{(2\pi)^3} D_{{1 \over 2} \bld k + \bld p} D_{{1 \over 2} \bld k - \bld p} {\cal G} (\bld k, \bld p)
\label{shape}
\end{equation}
Equation~(\ref{tevl}) governs the time evolution while Eq.~(\ref{shape}) governs the shape of
the power spectrum. (The  separation ansatz, of course, has the scaling freedom
$F\to \mu F, D\to (1/\mu)D$ which will change the right hand side of Eq.~(\ref{tevl}) to $-\mu F^2$ and the left hand side of Eq.~(\ref{shape})
to $\mu H_0^2 D_{\bf k}$. But, as to be expected, our results will be independent of $\mu$; so we have set it to unity). Our interest lies in analyzing the solutions of Eq.~(\ref{tevl}) subject to the initial conditions $F(a_i) =$ constant, $(dF/da)_i =0$ at some small enough $a=a_i$.

\begin{figure*}
\includegraphics[scale=0.5]{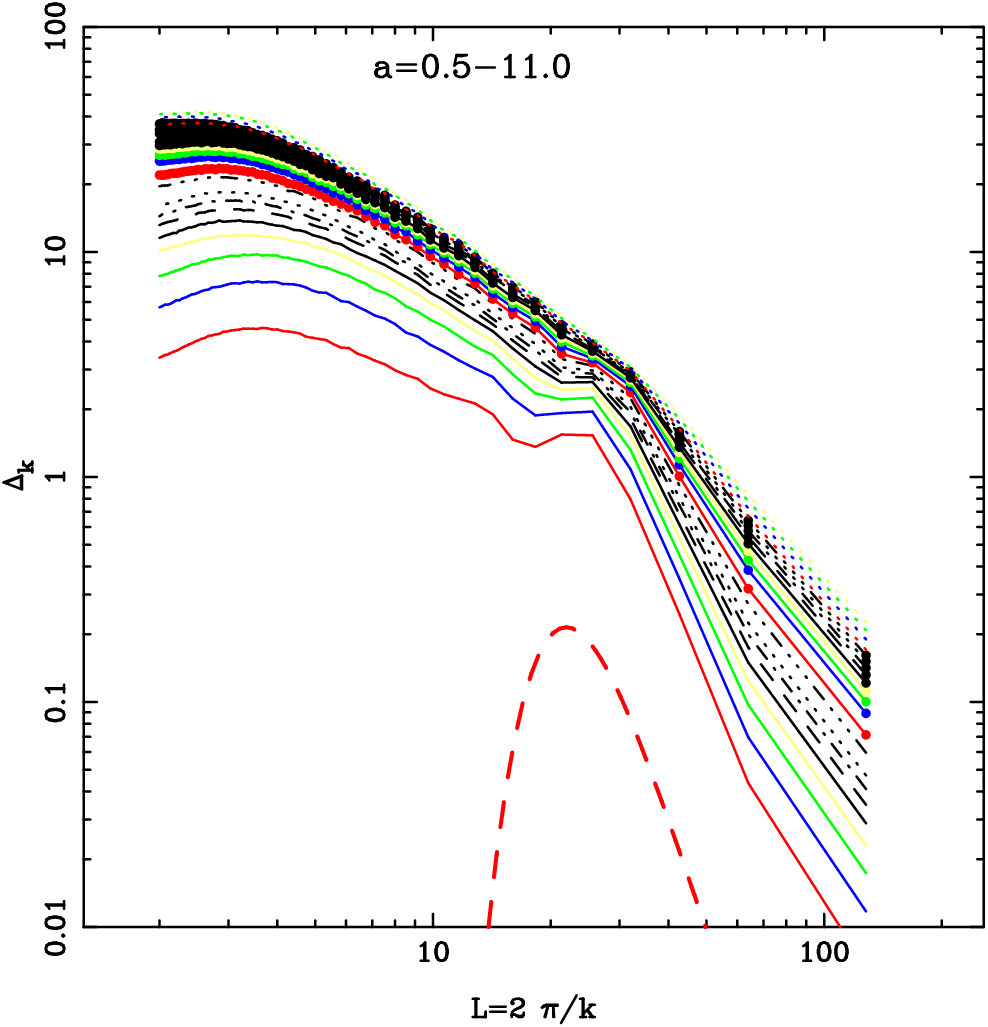}\ \includegraphics[scale=0.5]{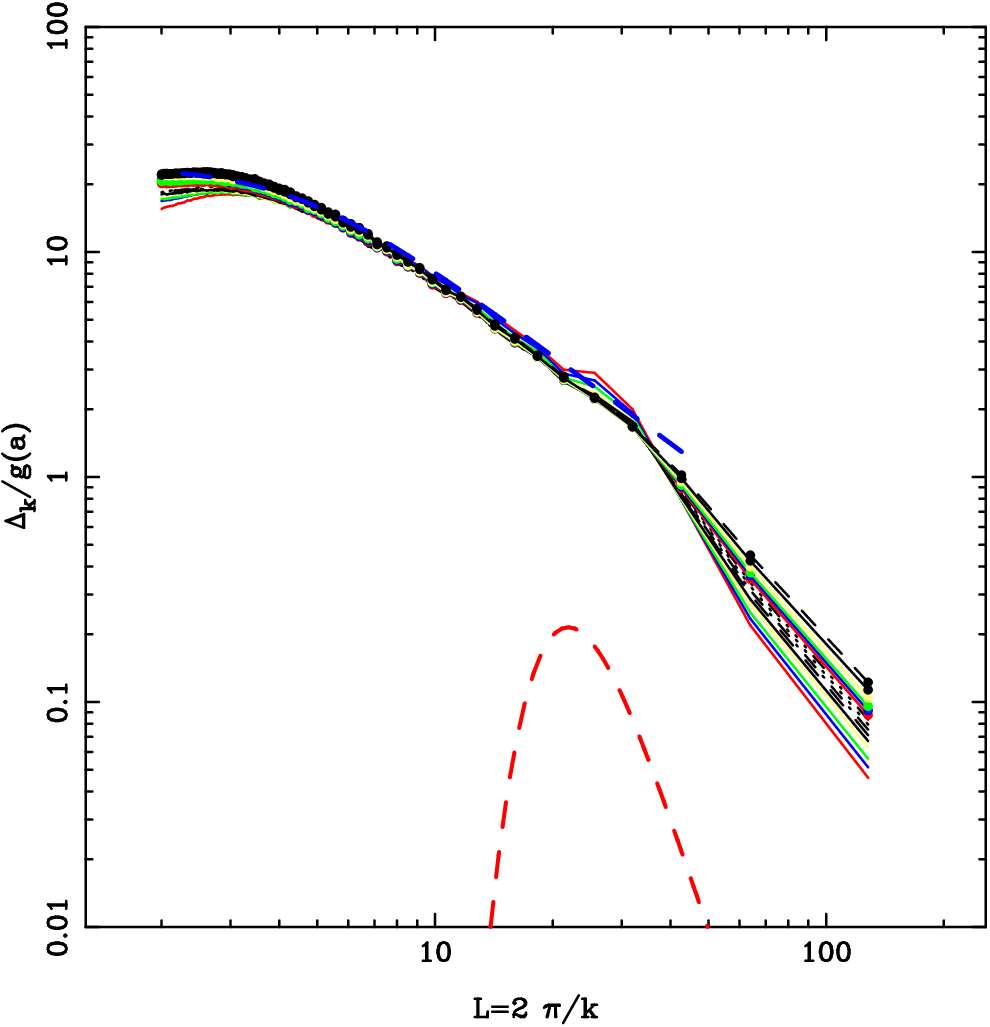}
\caption{Left panel: The results of the numerical simulation with an initial power spectrum which is 
a Gaussian peaked at $L=24$. The y-axis gives $\Delta_k$ where $\Delta_k^2=k^3 P(k)/2\pi^2$ is
the power per logarithmic band. The evolution generates a well known $k^4$ tail at large 
scales (see for example, \cite{jsbtp1})
and leads to cascading of power to small scales. Right panel: The simulation data is re-expressed
by factoring out the time evolution function $g(a)$ obtained by integrating Eq.~(\ref{gofa}). The fact
that the curves fall nearly on top of each other shows that the late time evolution is scale free
and described by the ansatz discussed in the text. The rescaled spectrum is very well described by
$P(k)\propto k^{-0.4}/(1+(k/k_0)^{2.6})$ which is shown by the completely overlapping, broken blue curve.}
\end{figure*}

Inspection shows that Eq.~(\ref{tevl}) has the exact solution $F(a) = (3/2) a^{-1}$.
This, of course, is a special solution and will not satisfy the relevant initial conditions. However,
Eq.~(\ref{tevl}) fortunately belongs to a  class of non-linear equations which can be mapped
to a homologous system. In such cases, the special power law solutions will arise as
the asymptotic limit. (The example well known to astronomers is that of isothermal sphere \cite{chandra}. Our
analysis below has a close parallel.)
To find the general behaviour of the solutions to Eq.~(\ref{tevl}), we will make the substitution
$F(a) = 
(3/2) a^{-1} g(a)$ and change the independent variable from $a$ to $q = \log a$.
Then Eq.~(\ref{tevl}) reduces to the form
 \begin{equation}
\frac{d^2 g}{d q^2} + \frac{1}{2} \frac{d g}{d q} + \frac{3}{2} 
g(g - 1) = 0
\label{gofa}
\end{equation} 
This represents a particle moving in a potential $V(g)= (1/2) g^3 - (3/4) g^2$ under 
friction.  For our initial conditions the motion will lead the ``particle'' to asymptotically come to rest at the stable minimum
at $g=1$ with damped oscillations. In other words, $F(a) \to (3/2) a^{-1}$ for large $a$
showing this is indeed the asymptotic solution. From the Poisson equation in Eq.(\ref{twnine}), it follows that $k^2\phi_\textbf{k}\propto (\delta_\textbf{k}/a)$ so that $\delta_\textbf{k}(a)\propto g(a)k^2D(\textbf{k})$ giving a direct physical meaning to the function $g(a)$. The asymptotic limit corresponds to a rather trivial case of $\delta_\textbf{k}$ becoming independent of time. What will be more interesting --- and accessible in simulations --- will be the approach to this asymptotic solution. To obtain this, we introduce
the variable $v=2(dg/dq)$ so that
our system reduces to the ``phase space'' equations
\begin{equation}
\dot v = - \frac{1}{2} v - 3  g(g - 1); \quad
\dot g = \frac{v}{2} 
\end{equation}
where the dot denotes differentiation with respect to $q$.  Dividing the first equation by
the second and changing variables to $u=(v+g)$, we get the first order form of the autonomous system to be 
\begin{equation} 
\frac{d u}{d g} = - \frac{6 g(g - 1)} {u - g}
\end{equation}
The critical points  of the system are at $(0,0) $ and $(1,1)$. Standard analysis shows that: (i) the first one is an unstable critical point and the second one is the stable critical point; (ii) for our initial
conditions the solution spirals around the stable critical point.

Figures 1(a) and 1(b) describe the solution in the $g-a$ and $u-g$ planes. The $g(a)$ curves clearly approach the asymptotic 
value of $g\approx 1$ with superposed oscillations. The different curves in Fig 1(a)
are for different initial values which arise from the scaling freedom mentioned earlier. (The thick red line correspond to the initial conditions used in the simulations described below.). Fig 1(b) shows this \textit{family} of solutions in the $u-g$ plane. As usual, in going from the second order equation for $g$ to the first order equation in $u-g$ plane, we map a family of solutions to a single curve \cite{chandra}. The stable critical point acts as an attractor. The solution
$g(a)$ describes the time evolution and solves the problem of determining asymptotic time evolution.

To test the correctness of these conclusions, we performed a high resolution simulation 
using the TreePM method
\cite{2002JApA...23..185B,2003NewA....8..665B} and its parallel version
\cite{2004astro.ph..5220Ra} with $128^3$ particles
on a $128^3$ grid. We used a cubic spline softened force with softening
length $\epsilon = 0.4$ to ensure collisionless evolution in the
simulations. We computed all relevant statistics at scales $L$
larger than $2 \epsilon$ to avoid errors due to the force softening.
Details about the code parameters can be found in
\cite{2003NewA....8..665B}.
The initial power spectrum $P(k)$ was chosen to be a Gaussian peaked at
the scale of $k_p = 2 \pi / L_p$ with $L_p = 24$ grid lengths and with a
standard deviation $\Delta k = 2 \pi / L_{box}$, where $L_{box} = 128$ grid lengths is
the size of one side of the simulation volume. The amplitude of the peak
was taken such that $\Delta_{lin}\left(k = k_p, a = 0.25\right)
= 1$.

The  late time evolution of the power spectrum (in terms of $\Delta_k^2\equiv k^3P(k)/2\pi^2$ where $P=|\delta_k|^2$ is the power spectrum of density fluctuations) obtained from the simulation is shown in Fig 2(a). In Fig 2(b), we have rescaled the $\Delta_k$, using the appropriate solution $g(a)$. The fact that the curves fall on top of each other shows that the late time evolution indeed scales as $g(a)$ within numerical accuracy.  A reasonably accurate fit for $g(a)$ used in this figure at late times  is given by
\begin{equation}
g(a)\propto a(1-0.3\ln a)
\end{equation}
The key point to note is that the asymptotic time evolution is essentially $\delta(a)\propto a$ except for a logarithmic correction, \textit{even in highly non-linear scales}. (This was first noticed from somewhat lower resolution simulations in \cite{jsbtp1}.). Since the evolution at \textit{linear} scales is always $\delta\propto a$,
 this allows for a form invariant evolution of power spectrum at all scales. Gravitational clustering evolves towards this asymptotic state. 

This behaviour is fairly generic and we have performed a series of simulations with 
different initial conditions to test this claim. Fig.~3 shows $\Delta_k$ scaled by $g(a)$ for a simulation 
 where we have initial power concentrated in two narrow windows in
$k$-space. In addition to power around $L_p=24$ grid lengths as in the previous case, we
added power at $k_1=2\pi/L_1$ ($L_1=8$ grid lengths) using a Gaussian with the same width
as before. Amplitude at $L_1$ was chosen to be five times higher than that at $L_p$, which means 
that $\Delta_{lin} (k_1,a=0.05)=1$. The fact that the curves fall on top of each other in this case too 
shows that the late time evolution indeed scales as $g(a)$, independent of initial conditions. 

  The form of the power spectrum in Fig.~2 is well approximated by 
\begin{equation}
P(k)\propto \frac{k^{-0.4}}{1+(k/k_0)^{2.6}}; \qquad \frac{2\pi}{k_0}\approx 4.5
\label{fit}
\end{equation}
This fit is shown by the broken blue line in the figure which completely overlaps with the data and is barely visible. (Note that this fit is applicable only at $L<L_p$ since the $k^4$ tail will dominate scales to the right of the initial peak; see the discussion in \cite {jsbtp1}). At non-linear scales $P(k)\propto k^{-3}$
making $\Delta_k$ flat, as seen in Fig.2. (This is \textit{not} a numerical artifact and we have sufficient dynamic range in the simulation to ascertain this.) At quasilinear scales
$P(k)\propto k^{-0.4}$. The power spectrum in Fig.~3 is also well-approximated by Eq.~(\ref{fit}), excepting for 
the fact that the scale $2 \pi / k_0 \approx 4.1$ in this case. A lower value for the scale is to be expected 
because of the additionally injected power at the lower scale $L_1$.

\begin{figure}
\includegraphics[scale=0.45]{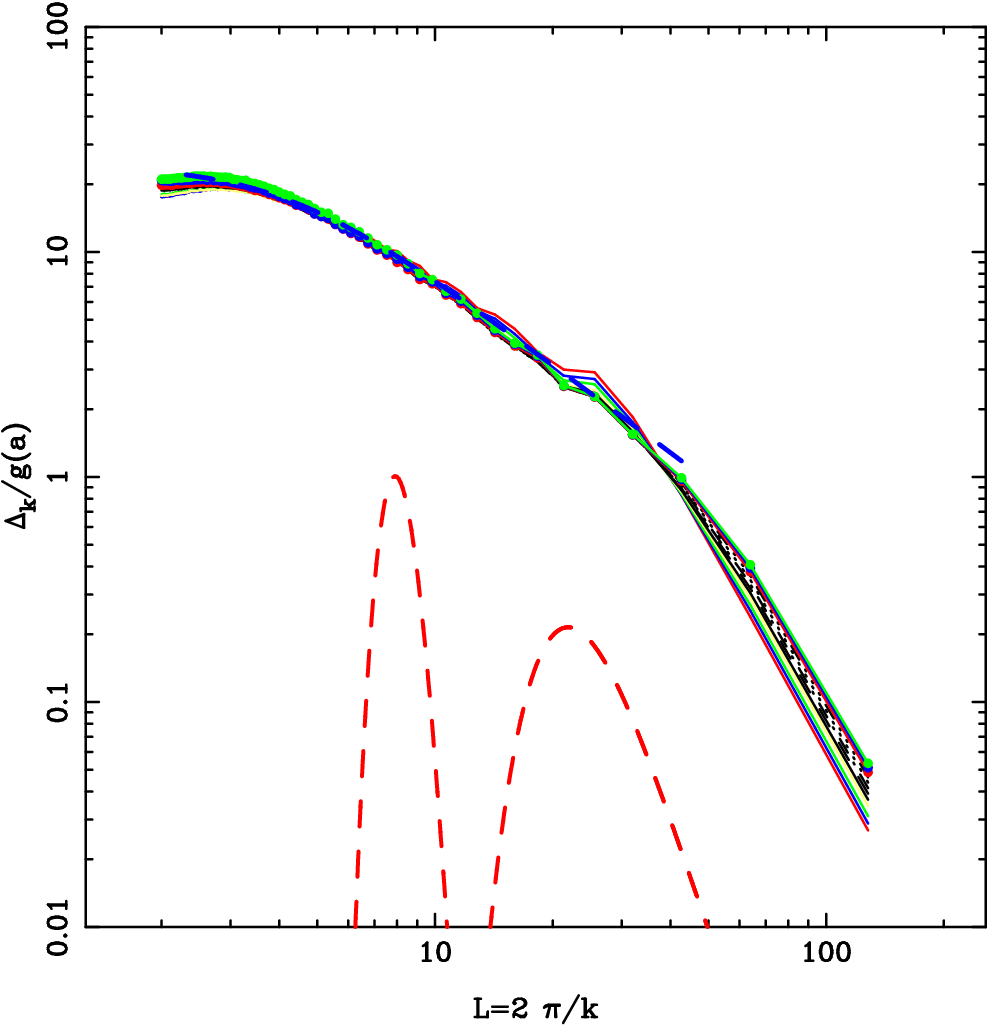}
\caption{The results of a numerical simulation with an initial power spectrum which 
has two Gaussians peaked at $L=24$ and $L=8$. The figure has been plotted 
by factoring out the time evolution function $g(a)$ (Eq.~(\ref{gofa})) from $\Delta_k$. The fact
that the curves fall nearly on top of each other in this case as well 
shows that the late time evolution is scale-free and is described by the ansatz discussed in the text,
{\sl independent of initial conditions}. The rescaled spectrum is once again described by
$P(k)\propto k^{-0.4}/(1+(k/k_0)^{2.6})$ which is shown by the completely overlapping, broken blue curve.}
\end{figure}

The effective index of the power spectrum varies between $-3$ and $-0.4$ in this range of scales for which 
the fitting function has been provided. To the lowest order of accuracy, the power spectrum at this range of scales is approximated by the mean index $n\approx -1$ with $P(k)\propto k^{-1}$. This index has a simple interpretation 
\cite{jsbtp1,klypin}. The ensemble-averaged gravitational potential energy of fluctuations per unit volume is given by
\begin{equation}
{\cal E}\propto \frac{1}{V}\int_V d^3\mathbf{x} d^3\mathbf{y}
\frac{\langle\delta(\mathbf{x})\delta(\mathbf{y})\rangle}{|\mathbf{x}-\mathbf{y}|}
\propto\int_0^\infty d^3r\frac{\xi(r)}{r}\propto\int_0^\infty \frac{dk}{k} kP(k)
\end{equation}
This energy reaches equipartition (i.e., contributes same amount per logarithmic band of scales) when
$P(k)\propto k^{-1}$. The same result holds for kinetic energy if the motion is dominated by scale invariant radial flows. Our result suggests that gravitational power transfer evolves towards this equipartition. In principle, this 
${\bf k}$ dependence of the power spectrum is determined by Eq.(\ref{shape}); but analytical understanding of this equation is more difficult since the solution depends on the phases of $\phi_{\bf k}$ which do not contribute to the power spectrum. This issue is under investigation.

We thank J.S. Bagla for useful discussions and comments on the draft. One of us (S.R.) thanks
Arnab Ray for useful comments.
Numerical experiments for this study were carried out at the cluster
computing facility in the Harish-Chandra Research Institute
(http://cluster.mri.ernet.in).


\begin{thebibliography}{99}

\bibitem{lssu} P.J.E. Peebles: \emph{Large Scale Structure of the Universe} (Princeton
University Press, New Jersey, 1980). 

\bibitem{gc1} T. Padmanabhan, astro-ph/0511536.   

%\bibitem{gc2} T. Padmanabhan, astro-ph/0512077.   
   
\bibitem{tppr}
T. Padmanabhan:  Physics Reports {\textbf 188}, 285 (1990); 
ApJ Supp. {\textbf 71}, 651 (1989) [astro-ph/0206131].

\bibitem{za} 
Ya.B. Zeldovich, A\&A \textbf{5}, 84,(1970);
S. N. Gurbatov et al, MNRAS \textbf{236}, 385 (1989);
T.G. Brainerd et al., ApJ \textbf{418}, 570 (1993);
S. Matarrese et al., MNRAS \textbf{259}, 437-452 (1992);
J.S. Bagla, T. Padmanabhan, MNRAS \textbf{ 266}, 227 (1994) [gr-qc/9304021];
T. Padmanabhan, S. Engineer, ApJ \textbf{493}, 509 (1998) [astro-ph/9704224];  
S. Engineer et al., MNRAS \textbf{314}, 279 (2000) [astro-ph/9812452]; 
for a recent review, see  T. Tatekawa [astro-ph/0412025].

\bibitem{pi} 
T. Buchert, MNRAS \textbf{267}, 811 (1994);
P. Valageas, A\&A \textbf{382}, 477 (2001); A\&A \textbf{379}, 8 (2001).

\bibitem{nsr}
A.J.S. Hamilton et al., ApJ \textbf{374}, L1 (1991);  
T. Padmanabhan et al., ApJ \textbf{466}, 604 (1996) [astro-ph/9506051];  
D. Munshi et al., MNRAS \textbf{290}, 193 (1997) [astro-ph/9606170];
J.S. Bagla et.al., ApJ \textbf{495}, 25 (1998) [astro-ph/9707330];
N. Kanekar et al., MNRAS \textbf{ 324}, 988 (2001) [astro-ph/0101562]; 
R. Nityananda, T. Padmanabhan, MNRAS \textbf{271}, 976 (1994) [gr-qc/9304022]; 
T. Padmanabhan, MNRAS \textbf{278}, L29 (1996) [astro-ph/9508124]. 

\bibitem{chandra}
S. Chandrasekhar: \emph{An Introduction to the Study of Stellar
Structure} (Dover, 1939).

\bibitem{2002JApA...23..185B}
J.S. Bagla, Journal of Astrophysics and Astronomy \textbf{23}, 185 (2002)
[astro-ph/9911025].

\bibitem{2003NewA....8..665B}
J.S. Bagla, S. Ray, New Astronomy \textbf{8}, 665 (2003).

\bibitem{2004astro.ph..5220Ra}
S. Ray, J.S. Bagla, astro-ph/0405220.

\bibitem{jsbtp1}
J.S. Bagla and T. Padmanabhan, MNRAS \textbf{286}, 1023 (1997).

\bibitem{klypin}
A.A. Klypin and A.L. Melott, ApJ \textbf{399}, 397 (1992).   
 
\end{thebibliography}
\end{document}